\documentclass[a4paper]{article}
\usepackage{graphicx} 
\usepackage{pdfpages}
\usepackage{placeins}
\usepackage{titling}    
\usepackage[a4paper, top=2cm, bottom=2.5cm, left=2cm, right=2.5cm]{geometry}
\usepackage{xcolor}
\usepackage{tikz}
\usepackage{braket}
\usepackage{amsmath}
\usepackage{titlesec}
\usepackage{setspace}   
\usepackage{float}
\usepackage{booktabs}
\usepackage{caption}
\usepackage{hyperref}
\usepackage{authblk}

\titleformat{\section}[block]{\normalfont\Large\bfseries\centering}{}{0pt}{}
\titleformat{\subsection}[block]{\normalfont\large\bfseries\centering}{}{0pt}{}

\title{Photonic Reservoir Engineering via 2D $\Lambda$-Type Atomic Arrays in Waveguide QED}
\author[1]{Thi Phuong Anh Nguyen}
\author[2]{Le Phuong Hoang}
\author[1,3]{Binh Xuan Cao\thanks{Corresponding author: binh.caoxuan@hust.edu.vn}}

\affil[1]{\small SquareLab, School of Mechanical Engineering, Hanoi University of Science and Technology, Hanoi, Vietnam}
\affil[2]{\small Institute of Science and Technology Austria, Klosterneuburg, Austria}
\affil[3]{\small Department of Mechatronics, School of Mechanical Engineering, Hanoi University of Science and Technology, Hanoi, Vietnam}
\date{}

\begin{document}
\maketitle
\begin{abstract}
Electromagnetically induced transparency (EIT) in $\Lambda$-type atomic systems underpins quantum technologies such as high-fidelity memory and nonlinear optics, but conventional setups face intrinsic limitations. Standard geometries of one-dimensional atomic chains coupled to waveguides allow only a single bright superradiant channel, while subradiant modes remain weakly accessible, limiting control over collective radiative behavior and dark-state pathways. This leads to unwanted inelastic processes, degrading memory fidelity and reducing nonlinear photon generation efficiency. Here, we propose two two-dimensional (2D) atomic lattice geometries coupled to a photonic crystal waveguide, namely Zigzag and Orthogonal structures. In the Zigzag model, engineered collective super- and subradiant modes produce a flattened EIT window, broadening the transmission bandwidth and suppressing unwanted scattering to enhance quantum memory fidelity. In the Orthogonal model, four-wave mixing (FWM) intensity is amplified by up to six orders of magnitude relative to a conventional one-dimensional $\Lambda$-type EIT chain with identical $\Gamma_{1D}$, $\Omega_c$, and probe intensity, with localized idler photons forming well-defined spectral modes. These results demonstrate a versatile route to engineer structured photonic reservoirs for on-demand photon generation, high-fidelity quantum storage, and enhanced nonlinear optical processes.
\end{abstract}
\vspace{1em}

\setstretch{2.0}  

\vspace{1em}

\section{I.   INTRODUCTION}
\label{sse: Intro}
\addcontentsline{toc}{section}{INTRODUCTION}
Interactions in $\Lambda$-type atomic waveguide quantum electrodynamics (wQED) have attracted considerable attention, as photon–photon and atom–photon interactions give rise to nontrivial phenomena underpinning modern quantum technologies. These include collective radiative effects such as superradiance and subradiance \cite{PhysRevA.110.053716}, bound states in the continuum (BIC) \cite{PhysRevA.104.013705, PhysRevA.97.043848}, and non-Markovian retardation effects \cite{PhysRevLett.131.193603}. All of these phenomena depend sensitively on the spatial arrangement of atoms, highlighting the central role of phase relations and interference in controlling system dynamics.\\
A distinctive feature of $\Lambda$-type atoms is their ability to exhibit electromagnetically induced transparency (EIT), where Fano interference between the $|g\rangle \leftrightarrow |e\rangle$ and $|s\rangle \leftrightarrow |e\rangle$ transitions enables the lossless propagation of dark-state polaritons \cite{RevModPhys.77.633}. Under adiabatic conditions, optical fields can be coherently mapped into atomic spin excitations, forming the basis for quantum memories \cite{Baba2008}. Moreover, when atoms are prepared in the laser-dressed dark-state configuration, the linear susceptibility vanishes while the nonlinear susceptibility is enhanced, making EIT an ideal medium for nonlinear processes such as four-wave mixing (FWM) \cite{RevModPhys.77.633, Offer2018}.\\
Despite these advantages, conventional EIT-based systems face significant limitations. In quantum storage, input pulses must be carefully shaped to fit within the narrow transparency window to avoid distortion \cite{Ding2013, PhysRevA.102.063720}. Stochastic inelastic processes such as Raman scattering and FWM introduce noise and dephasing that degrade storage fidelity. In nonlinear regimes, imperfect mode and phase matching broaden the inelastic spectral distribution, limiting photon coherence \cite{PhysRevA.110.063723, PhysRevApplied.14.024013, Novikov2016}. Physically, these limitations arise from the broadband nature of the inelastic scattering reservoir \cite{RevModPhys.77.633, Prajapati:17}. A promising strategy is to confine inelastic scattering into discrete photonic modes, effectively structuring the reservoir to support coherent nonlinear interactions.\\
Photonic crystal waveguides (PCWs) offer a versatile platform to implement this idea. By engineering atomic positions and coherent exchange strengths, both coherent and dissipative atom–photon interactions can be controlled. Motivated by this, we propose a system combining $\Lambda$-type atomic ensembles with a one-dimensional PCW, where atoms are arranged in two-dimensional zigzag and orthogonal lattices (Sec.~II). Variations in atomic height above the waveguide modify individual decay rates $\Gamma_{1D}$ \cite{Lee2025ExcitationLocalization, pnas.1603788113}, enabling structured dissipation and tunable interference. This geometric control facilitates a transition from conventional single-photon transmission to correlated multi-photon dynamics, where strong nonlinear optical responses emerge.\\
In the following sections, we develop the theoretical framework of this 2D-$\Lambda$-type wQED system, derive photon scattering properties using input–output theory, and analyze the emergence of transmission passbands and nonlinear photon correlations in these structured lattices.
\section {II.   THEORETICAL MODEL}
\label{sec:Theory-model}
\addcontentsline{toc}{section}{A. THEORETICAL DESCRIPTION OF PHOTONS SCATTERING INSIDE THE SYSTEM}
\subsection {Hamiltonian of the PCW with a 2D-atomic lattice}
\addcontentsline{toc}{subsection}{A. Hamiltonian of the PCW with a 2D-atomic lattice}
In free space, atoms couple weakly to light, making it difficult to engineer strong and controllable light–matter interactions \cite{Douglas2015}. By contrast, a one-dimensional (1D) waveguide enhances control over the atomic response to electromagnetic fields \cite{tevcer2024strongly}. In such a system, light is confined to a single propagation axis, facilitating stronger and more controllable atom–photon interactions. Along the waveguide, atoms arranged in a periodic lattice produce Bragg interference between counter-propagating fields. It is this atomic lattice periodicity, rather than solely the inherent photonic crystal band structure, that modifies the local photonic density of states (LDOS), leading to frequency-dependent enhancement (superradiance) or suppression (subradiance) of emission \cite{PhysRevA.110.053716, PhysRevB.96.144201}. As a result, atoms trapped near the waveguide form a controllable quantum material, where both linear and nonlinear optical responses can be tuned by geometrical characteristics.\\
We consider two-dimensional (2D) lattices of $\Lambda$-type atoms positioned near the lower band edge of a photonic crystal waveguide (PCW) [see Figs. \ref{Fig. Zigzag model} and \ref{Fig. Orthogonal model}]. This configuration enables independent control over both the individual (self-) and collective dissipative decay rates into the guided mode \cite{Lee2025ExcitationLocalization, pnas.1603788113}, thereby allowing precise control of light transmission through a structured photonic reservoir in the frequency domain. The atomic states $\ket{g}$, $\ket{s}$, and $\ket{e}$ correspond to the ground, metastable, and excited states, respectively, with all atoms (represented by blue circles) initially prepared in the ground state $\ket{g}$. Our two proposed models contain a PCW structure with different $\Lambda$-type atomic lattice geometries, as illustrated in Fig. \ref{Fig. Zigzag model} and Fig. \ref{Fig. Orthogonal model}. We assume that the resonance frequency $\omega_0$ of the transition $\ket{g}\leftrightarrow\ket{e}$, with wave vector $k_0$, lies inside the TE band gap and close to its lower edge, with detuning $\delta = \omega_0 - \omega_b$. In this regime, due to the van Hove singularity in the density of states, the transition $\ket{g}\leftrightarrow\ket{e}$ is coupled dominantly to the PCW mode near the lower band edge.\\
Within the effective-mass approximation, the dispersion relation of the TE mode near the band edge can be expressed as
$\omega_k \approx \omega_b\left[1 - \alpha \frac{(k - k_b)^2}{k_b^2}\right]$,
where $k_b = \pi/a$ is the wave vector at the band edge and $\alpha$ characterizes the band curvature. When an atom couples to this TE mode at a frequency inside its band gap, it cannot emit a propagating photon but instead seeds an exponentially decaying localized photonic cloud. This localized field acts as an effective cavity-like mode, mediating long-range excitation exchange between atoms through virtual photons. In contrast, the TM mode does not exhibit a band gap near the atomic resonance frequency $\omega_0$ and thus serves as a propagating channel for probing transmission and long-range interactions, as the TE mode is fully band-gapped at near $\omega_0$. In our configuration, the transition $\ket{g}\leftrightarrow\ket{e}$ couples to the TM mode via evanescent fields, while the transition $\ket{s}\leftrightarrow\ket{e}$ is driven by a classical control field with Rabi frequency $\Omega_c$. Under a weak probe detuned by $\Delta$ from resonance, and within the rotating-wave approximation, the atomic chain coupled to the PCW can be described by the effective non-Hermitian Hamiltonian given in Eq.~\ref{Eq. Hamiltonian} \cite{Caneva_2015, PhysRevA.98.023814}.
\begin{align}
H_{\text{non}} &= -\sum_{j=1}^{n}[
   \big(\Delta\omega + i\Gamma_e/2\big)\,\sigma_{ee}^j
   + (\Delta\omega - \Delta_c)\,\sigma_{ss}^j 
+ \Omega_c \big(\sigma_{es}^j + \text{H.c.}\big)] +  H_{\Gamma_{1D}} +  H_{\mathcal{J}}.
\label{Eq. Hamiltonian}
\end{align}
where $\Delta\omega = \omega_{in} - \omega_0$ is the detuning between the frequency $\omega_{in}$ of the incident field with wave vector $k_{in}$ and the atomic resonance frequency $\omega_0$. $\Gamma_e$ represents the decay rate of the state $\ket{e}$ into free space and $z_j$ is the position of the $j$th atom. $\Delta_c = \omega_c - \omega_{es}$ is the detuning between the frequency $\omega_c$ of the classical control field and the frequency $\omega_{es}$ of the atomic transition $\ket{e} \leftrightarrow \ket{s}$.\\
The last two terms, $H_{\Gamma_{1D}}$ and $H_{\mathcal{J}}$ (detailed in Appendix \ref{sec: APPENDIX A}2), describe the collective atom--atom interactions mediated by the guided modes of the photonic crystal waveguide. Specifically, $H_{\mathcal{J}}$ accounts for the coherent dipole--dipole exchange interactions arising from virtual photon exchange, while $H_{\Gamma_{1D}}$ represents the collective non-Hermitian dissipation into the one-dimensional guided continuum. Their explicit forms depend on the lattice geometry and atomic configuration, and will be discussed in detail in Sec. 2\ref{sec: 2B} for each proposed model.\\
\subsection{B. 2D $\Lambda$-type atomic lattices: for quantum memory and four-wave mixing}
\label{sec: 2B}
\begin{figure}[h]
    \centering
    \includegraphics[width=0.7\textwidth]{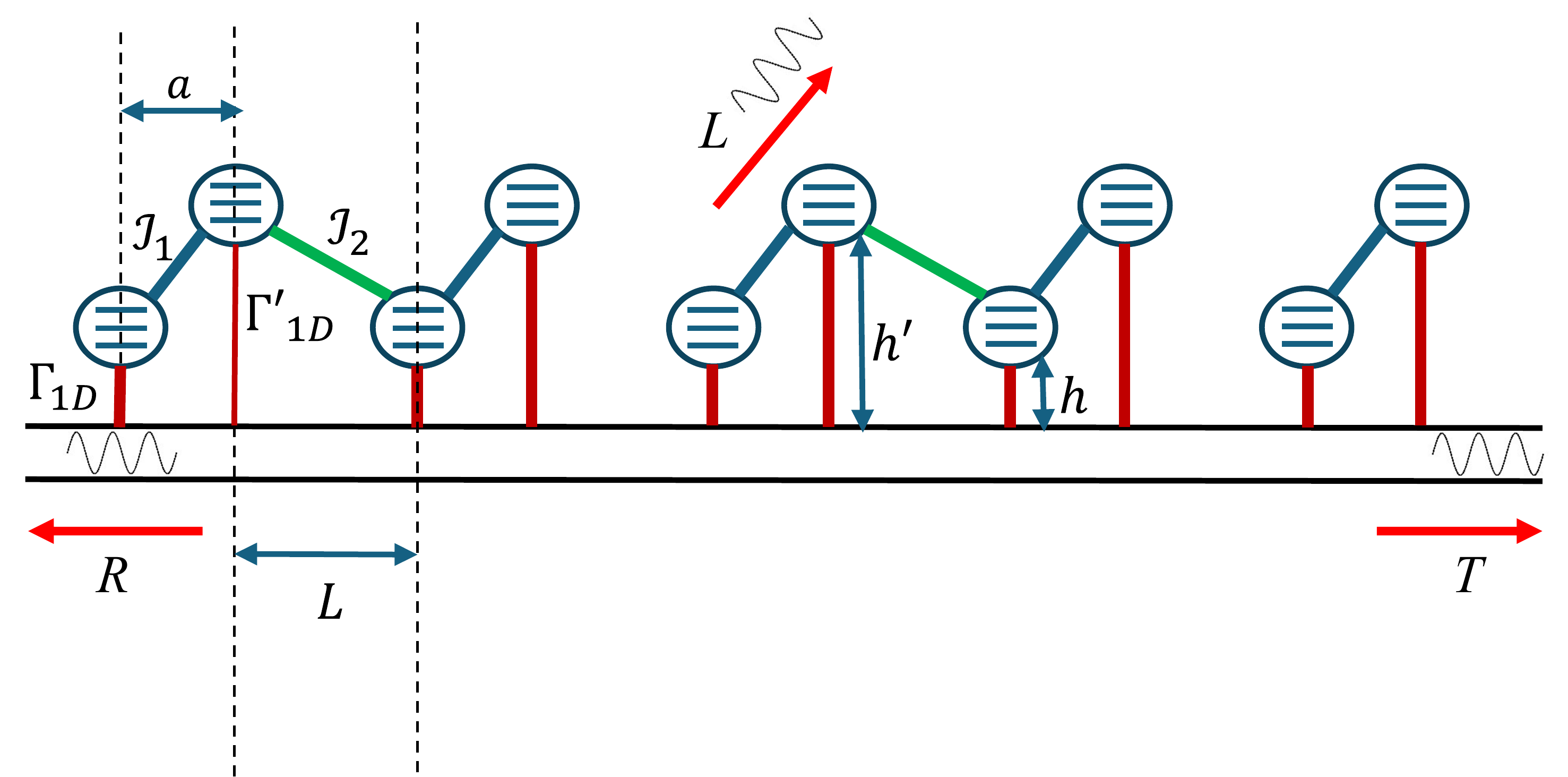}
    \caption{Zigzag model: Schematic of an atomic chain of ensembles coupled to a photonic crystal waveguide, forming a multi-photonic structured reservoir for quantum-memory applications.}
    \label{Fig. Zigzag model}
\end{figure}

\begin{figure}[h]
\centering
\includegraphics[width=0.7\textwidth]{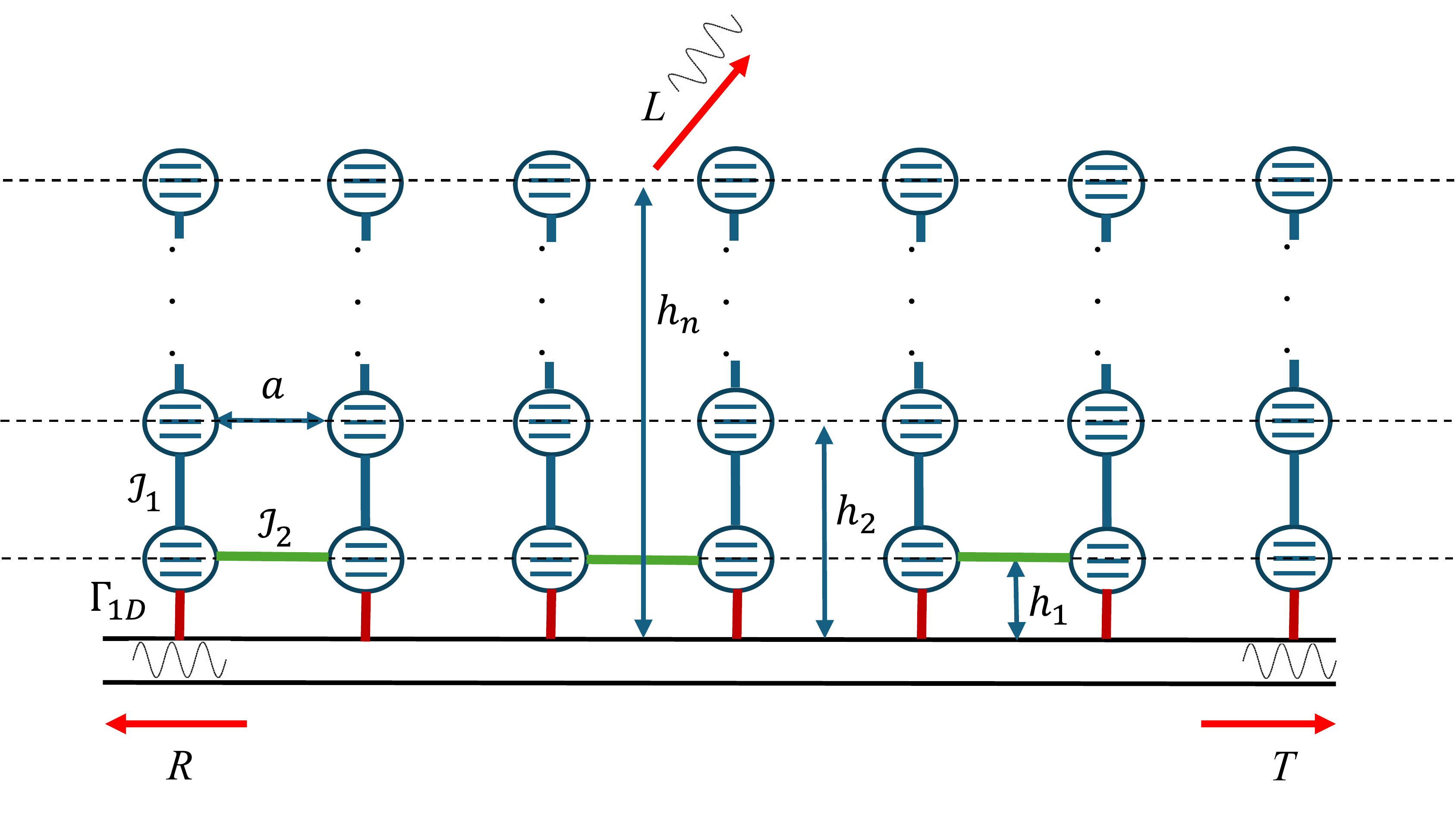}
\caption{Orthogonal model: Schematic of an atomic chain of ensembles coupled to a photonic crystal waveguide, forming a multi-photonic structured reservoir for nonlinear light--matter interactions.}
\label{Fig. Orthogonal model}
\end{figure}

\addcontentsline{toc}{subsection}{B. Proposed models for quantum memory and four-wave mixing processes}
\textbf{Model 1: Zigzag 2D-lattice structure.}\\
To realize quantum memory functionalities, we propose a zigzag two-dimensional (2D) lattice of $\Lambda$-type atoms coupled to a photonic crystal waveguide, as illustrated in Fig. \ref{Fig. Zigzag model}. The lattice consists of periodic clusters, each formed by two unit cells. Within each cell, two atoms are positioned at different vertical heights relative to the waveguide and separated by an intra-cell distance $a$, while adjacent cells in a cluster are separated by a distance $L$.
\noindent
\noindent
The zigzag geometry is designed to suppress Bragg interference and flatten the electromagnetically induced transparency (EIT) window. This is achieved by choosing the intra-cell and inter-cell phase delays as $\alpha = n\pi/2$ and $\beta = m\pi/2$ ($m>n$), which cancel Bragg reflections and enhance the EIT bandwidth. Due to their different vertical positions, the two atoms in each cell couple to the guided mode with distinct decay rates \cite{Lee2025ExcitationLocalization}. For simplicity, we consider one atom with a finite decay rate $\Gamma_{1D}$ and the other effectively decoupled from the waveguide. Atoms within the same cell interact via an intra-cell coupling $J_1$, while neighboring cells within a cluster are coupled by $J_2$ (Fig. \ref{Fig. Zigzag model}). We focus on the dissipation-dominated regime $\Gamma_{1D} \gg \mathcal{J}$, where coherent exchange through the guided mode can be neglected \cite{PhysRevA.110.053716}. This configuration yields a broad and robust EIT window suitable for quantum memory applications.\\
\textbf{Model 2: Orthogonal 2D-lattice structure.}\\
To enhance nonlinear light--matter interactions, we consider an orthogonal 2D lattice configuration, shown in Fig. \ref{Fig. Orthogonal model}. The lattice is divided into clusters, each consisting of two unit cells separated by a distance $a$. Each cell contains $N$ $\Lambda$-type atoms located at the same horizontal position but at different vertical heights relative to the waveguide.
\noindent
\noindent
In this model, only the atom closest to the waveguide in each cell has a significant decay rate $\Gamma_{1D}$, while the remaining atoms are effectively decoupled from the guided mode. Within each cell, atoms are sequentially coupled via coherent exchange with strength $\mathcal{J}_1$ \cite{PhysRevA.97.043848}, and the two cells in a cluster are connected through a coupling $\mathcal{J}_2$. There is no direct coupling between different clusters, resulting in a localized and hierarchical interaction structure. In contrast to Model~1, which suppresses Bragg interference, this model operates in the constructive interference regime. The lattice spacing is chosen as $k_0 a = \pi/2$, ensuring constructive interference of guided fields emitted by adjacent cells. This enhances collective dissipation and promotes strong optical nonlinearities, enabling multi-EIT window formation and efficient four-wave mixing within each cluster.
\section {III. Results of localizing photonic mode for quantum memory}
\label{sec: Results of Zigzag modelmodel}
\begin{figure}[h]
    \centering
    \includegraphics[width=\textwidth]{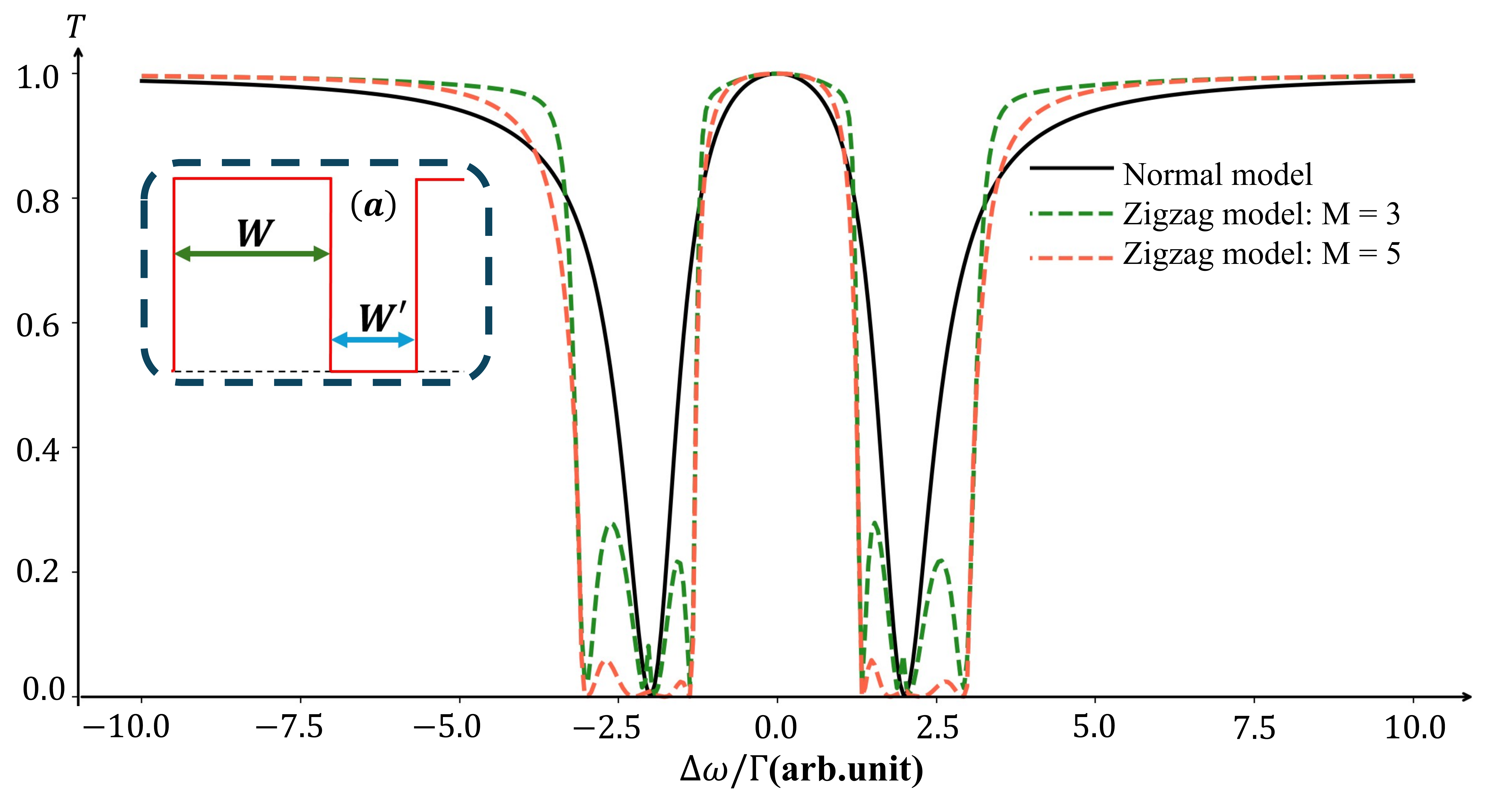}
    \caption{One-photon transmission spectra of $\Lambda$-type atomic schemes as a function of the probe detuning $\Delta\omega$. The black solid line corresponds to the conventional EIT model, while the green and orange dashed lines represent the Zigzag models with $M=3$ and $M=5$, respectively. The simulation parameters: $\Gamma_{1D} = \Gamma$, $\Gamma_e = 0.1\Gamma$, $\Omega_c = 2\Gamma$, $J_1 = 0.2\Gamma$, $J_2 = 1.6\Gamma$, $\Delta_c = 0$, $k_0 d = \pi/2$, and $a = d$. (a), Engineered EIT-band profile for quantum memory, where $W$ denotes the bandwidth of the EIT transparency window and $W'$ indicates the bandwidth of the adjacent absorption band.}
    \label{Fig. One-photon transmission through Zigzag model}
\end{figure}

\addcontentsline{toc}{section}{Results of localizing photonic mode for quantum memory}
\addcontentsline{toc}{subsection}{A. One-photon transmission spectrum}
In EIT-based quantum memories, optical storage is realized by adiabatically reducing the control field, mapping a photonic excitation onto a long-lived collective spin state. Although adiabaticity enables reversible storage, it is constrained by two main limitations. First, the narrow Lorentzian transparency window leads to uneven transmission of different spectral components, resulting in dispersion, distortion, and energy loss due to excess phase fluctuations in the spectral wings \cite{RevModPhys.89.015006}. Second, inelastic scattering processes, including spontaneous Raman scattering and four-wave mixing, generate unwanted Stokes and anti-Stokes photons that degrade the coherence of the stored state.\\
Fig. \ref{Fig. One-photon transmission through Zigzag model} compares the one-photon transmission spectra of the Zigzag model for \(M=3\) and \(M=5\) with a conventional two-atom PCW-based EIT system (black line). Unlike the standard Lorentzian transparency window, the Zigzag configuration exhibits a flatter transmission peak bounded by two pronounced absorption bands. These bands arise from subradiant modes induced by collective dissipation into the waveguide, which spectrally isolate the EIT window. Within the window, superradiant modes dominate and compensate transparency loss away from resonance. Importantly, this spectral structuring stabilizes the adiabatic storage process more effectively than simply increasing the control-field Rabi frequency \(\Omega_c\), which broadens the bandwidth but cannot prevent non-adiabatic distortions when the condition \(\dot{\Omega}_c \ll \Omega_c^2/\Gamma\) is violated~\cite{PhysRevA.102.063720}.\\
The near-rectangular EIT profile produced by the zigzag model [Fig. \ref{Fig. One-photon transmission through Zigzag model}(a)] addresses loss constraints in quantum memory operation. The engineered transparency window exhibits nearly uniform transmission over a broadened bandwidth, relaxing pulse-shaping requirements while preserving adiabatic propagation. The sharp absorption edges arise from collective superradiant and subradiant modes, which rapidly dissipate bright excitations and thereby spectrally isolate the coherent transmission channel from inelastic scattering pathways. Importantly, the flattening of the EIT window occurs without introducing anomalous dispersion, as confirmed by the smooth and approximately linear behavior of $\chi'$ within the transparency region. The zero crossings of $\chi'$ at the EIT boundaries mark the energetic separation between the quasi-dark-state manifold and the lossy bright collective modes, enforcing an effectively dynamical isolation of the dark-state polariton. While superradiant modes shape the steep transmission edges within the detuning range $[-\Gamma,\Gamma]$, subradiant modes broaden the surrounding absorption bands, together forming the characteristic near-rectangular spectrum.
\noindent
\noindent
Small oscillatory ripples inside the absorption bands originate from interference between coherently and incoherently coupled atomic cells. These features are progressively suppressed as the number of cells increases, indicating effective ensemble averaging and enhanced spectral isolation. As summarized in Table~\ref{tab:WWprime}, increasing the inter-cell coupling $J_2$ (with fixed $J_1$) narrows the EIT window while broadening the absorption bands, thereby strengthening the spectral separation between transmission and absorption regions and yielding an isolated quasi-dark-state manifold for high-fidelity quantum memory. 
\noindent
\begin{table}[h]
\caption{EIT and absorption band widths for different $J_1$ and $J_2$ values.}
\label{tab:WWprime}
\centering
\setlength{\tabcolsep}{10pt}
\begin{tabular}{ccccc}
\toprule
\textbf{No.} & $\mathbf{J}_1/\Gamma$ & $\mathbf{J}_2/\Gamma$ & $\mathbf{W}'/\Gamma$ & $\mathbf{W}/\Gamma$ \\
\midrule
1 & 0.2 & 1.6 & 1.66 & 2.69 \\
2 & 0.2 & 2.0 & 2.06 & 2.52 \\
3 & 0.2 & 2.4 & 2.46 & 2.24 \\
4 & 0.2 & 2.8 & 2.84 & 2.13 \\
5 & 0.2 & 3.2 & 3.23 & 1.95 \\
6 & 0.2 & 3.6 & 3.59 & 1.84 \\
7 & 0.2 & 4.0 & 4.10 & 1.68 \\
\bottomrule
\end{tabular}
\end{table}
\FloatBarrier
\noindent
\addcontentsline{toc}{subsection}{B. Inelastic scattering spectrum in frequency space for two photons}
To evaluate the system beyond single-photon transmission, we consider two-photon scattering. In a conventional system, inelastic scattering is weak but broadly distributed over $-5\Gamma$ to $5\Gamma$, making filtering ineffective.
\noindent
\begin{figure}[H]
    \centering
    \includegraphics[width=\textwidth]{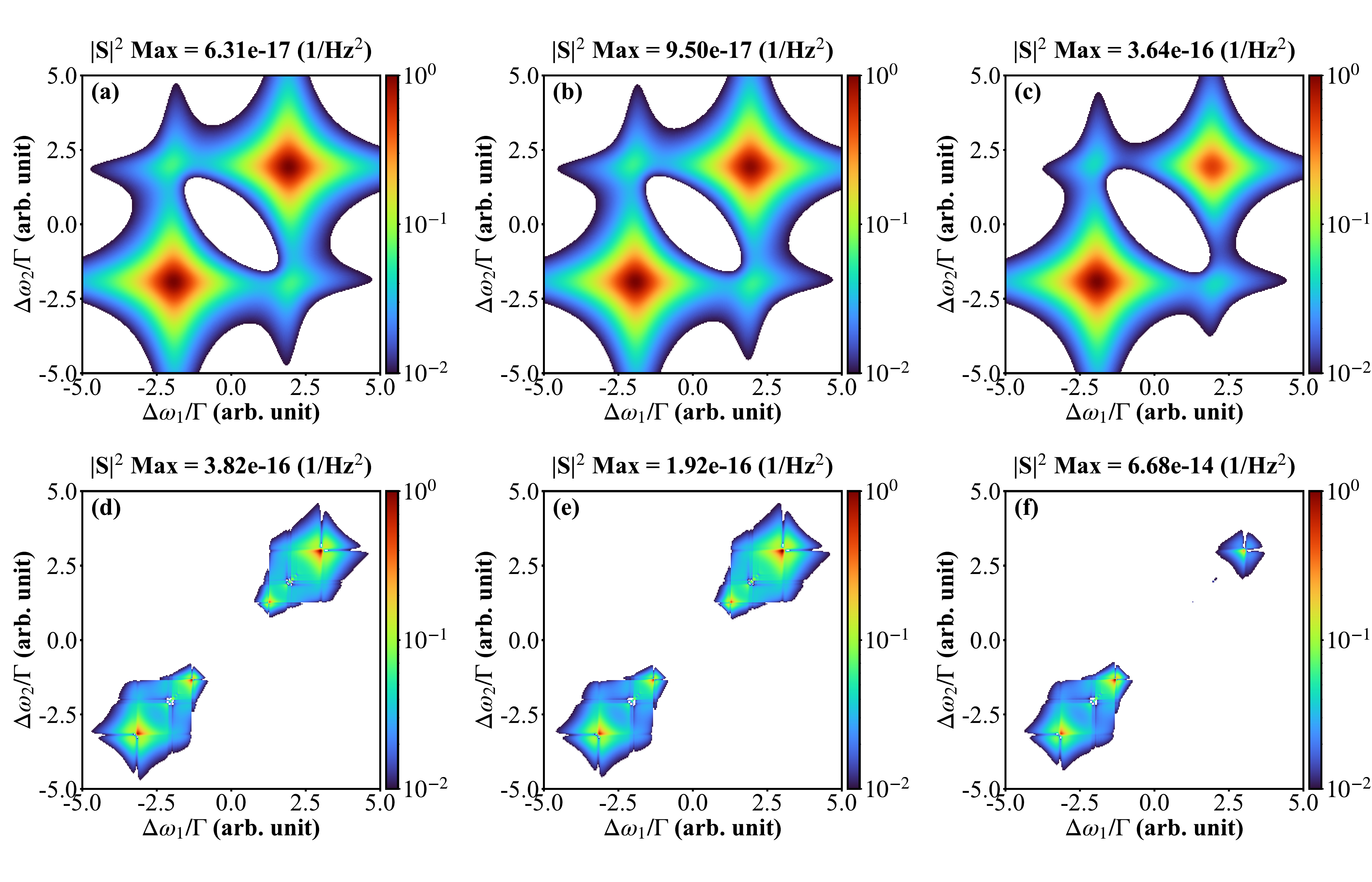}
    \caption{Inelastic scattering spectra for two incident photons as a function of probe detuning $\Delta\omega$. Panels (a)–(c): conventional EIT model; (d)–(f): Zigzag model. Two-photon inputs with detunings $\Delta\omega = 0$ [(a), (d)], $0.5\Gamma$ [(b), (e)], and $\Gamma$ [(c), (f)]. The simulation parameters: $\Gamma_{1D} = \Gamma$, $\Gamma_e = 0.1\Gamma$, $\Omega_c = 2\Gamma$, $J_1 = 0.2\Gamma$, $J_2 = 1.6\Gamma$, $\Delta_c = 0$, $k_0 d = \pi/2$, and $a = d$.}
    \label{Fig. Inelastic scattering inside Zigzag model}
\end{figure}
\noindent
In contrast, the Zigzag model ($M = 5$) confines inelastic scattering to two symmetric regions around the resonance, with peaks shifted by approximately $\Gamma$ from the positions $\pm 2\Gamma$ set by the control field. These positions correspond to the dressed bright states of the system, resulting from Autler–Townes splitting under the control field. Specifically, the peak frequencies follow
$\omega = \pm \frac{\Omega^2}{2\Gamma}$, where $\Omega$ is the Rabi frequency of the control field, reflecting the energy splitting of the laser-dressed bright states. Their magnitudes remain low ($\sim 10^{-16}\ \mathrm{Hz}^{-2}$), comparable to the conventional EIT case, but occur outside the EIT bandwidth, minimizing their effect on coherent transmission.\\
This localization of inelastic photons allows efficient filtering and reduces noise that could degrade the stored quantum state. Together with the flat-top EIT window and broadened absorption bands, the Zigzag model significantly enhances transmission fidelity and the performance of quantum memory. These results demonstrate that engineering both the transparency window and the absorption bands provides a practical route toward high-fidelity EIT-based quantum memory.   
\section {IV. Results of localizing photonic mode for four-wave mixing}
\label{sec: Results of Orthogonal modelmodel}
\begin{figure}[H]
    \centering
    \includegraphics[width=1.0\textwidth]{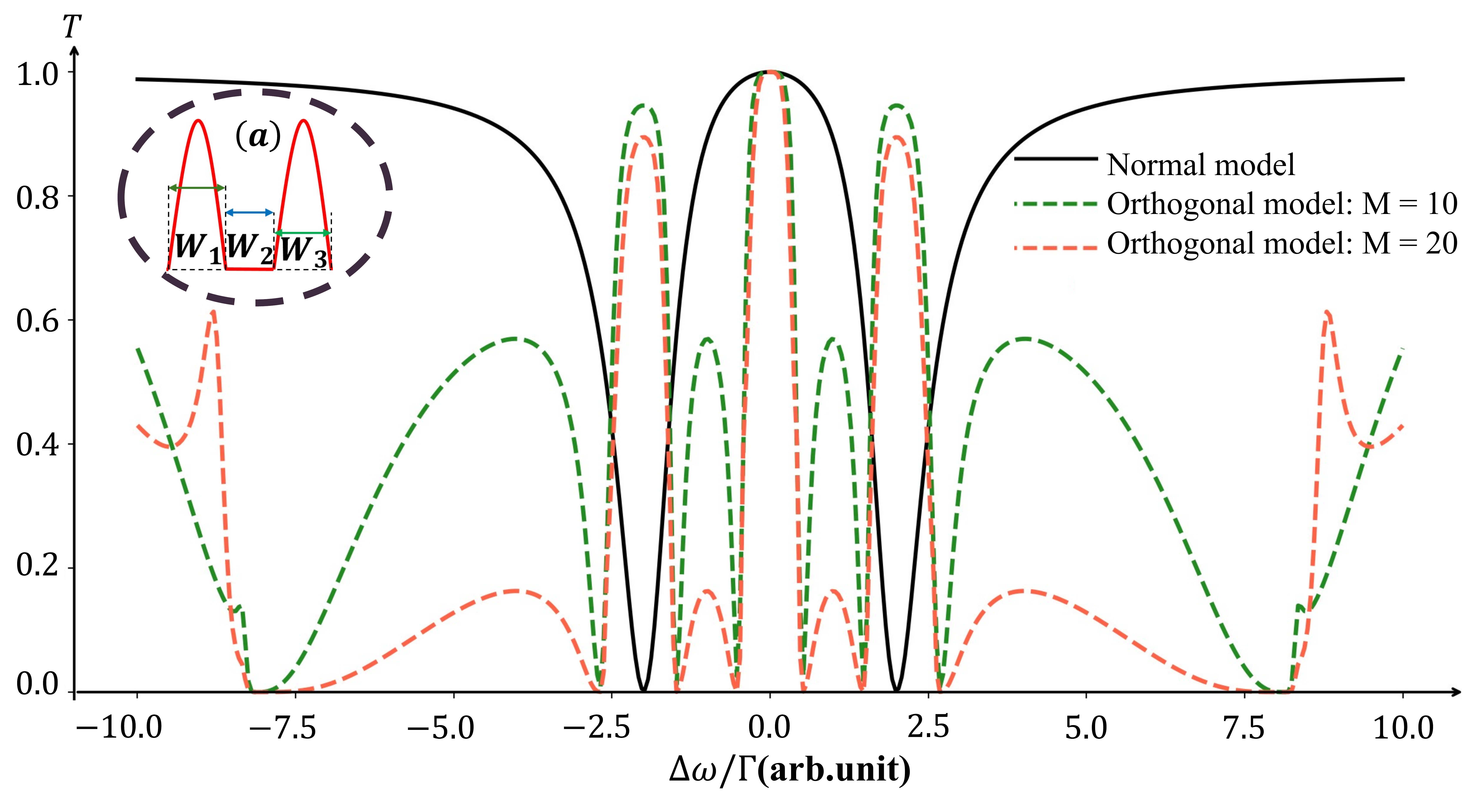}
    \caption{One-photon transmission spectra of $\Lambda$-type atomic schemes as a function of the probe detuning $\Delta\omega$. The black solid line corresponds to the conventional EIT model, while the green and orange dashed lines represent the Orthogonal models with two atoms per unit cell $N = 2$, and with $M=10$ and $M=20$, respectively. The simulation parameters are $\Gamma_{1D} = \Gamma$, $\Gamma_e = 0.1\Gamma$, $\Omega_c = 2\Gamma$, $J_1 = 3\Gamma$, $J_2 = 6\Gamma$, $\Delta_c = 0$, $k_0 d = \pi/2$, and $a = d$. (a) Engineered transmission profile for the four-wave mixing (FWM) response, where $W_1$ and $W_3$ denote the bandwidths of the additional and main EIT transparency windows, respectively, and $W_2$ corresponds to the absorption band separating these two transparency windows.}
    \label{Fig. One photon transmission through Orthogonal model}
\end{figure}

\addcontentsline{toc}{section}{Results of localizing photonic mode for four-wave mixing}
\addcontentsline{toc}{subsection}{A. One-photon transmission spectrum}
Laser dressing enables the EIT medium to operate as an effective nonlinear platform, where the transparency window not only suppresses loss but also spectrally structures the surrounding absorption bands. In contrast to conventional schemes that avoid absorption altogether, our orthogonal model exploits these absorption features to enhance nonlinear wave mixing. The orthogonal atomic arrangement suppresses destructive Bragg interference while generating multiple, well-resolved transmission resonances. Under a strong control field, laser dressing produces distinct side absorption bands around the central EIT window, as shown in Fig. \ref{Fig. One photon transmission through Orthogonal model}. These bands exhibit enhanced optical density and steep dispersion, making them optimal for nonlinear interactions. In this design, the absorption bands are intentionally kept separated by transmission windows, as illustrated in Fig. \ref{Fig. One photon transmission through Orthogonal model}(a). This spectral separation preserves frequency selectivity and prevents excessive broadening when multiple nonlinear channels are accessed simultaneously. As a result, the nonlinear response is maximized near the absorption resonances without compromising spectral control. By tuning the pump field to these peaks, photons couple resonantly through EIT-assisted absorption channels, giving rise to frequency-resolved inelastic scattering pathways.
\noindent
\noindent
Under weak-field excitation, the single-photon transmission spectrum for \(N=2\) atoms per unit cell exhibits three transparency windows separated by absorption bands. A central peak at zero detuning and two symmetric side peaks at \(\pm 2\Gamma\) with transmission exceeding 0.9 originate from superradiant collective dissipation in the waveguide. The absorption bands at $\Delta\omega/\Gamma \approx \pm 0.5,\,1.5,$ and $2.5$ (see Fig. \ref{Fig. One photon transmission through Orthogonal model}) are instead associated with subradiant collective modes of the unit cell. These features correspond to poles of the retarded Green’s function, as discussed in Appendix \ref{sec: APPENDIX A}1, where destructive collective emission suppresses transmission and induces strong dispersive responses. The positions of these subradiant bands are set by intra-cell interactions and the inter-cell phase shift \(\pi/2\), and coincide with frequencies where two-photon and four-wave mixing processes are enhanced. The number of transmission peaks scales with the number of atoms per unit cell, while the chosen inter-cell phase prevents Bragg cancellation and preserves constructive interference among the peaks. Residual ripple-like features within the absorption bands originate from resonant–anti-resonant interference and are progressively suppressed as the number of cells along the PCW increases, indicating enhanced collective coupling while maintaining sharp spectral selectivity.
\addcontentsline{toc}{subsection}{B. Inelastic scattering spectrum in frequency space for two photons}
\noindent
\begin{figure}[H]
    \centering
    \includegraphics[width=1.0\textwidth]{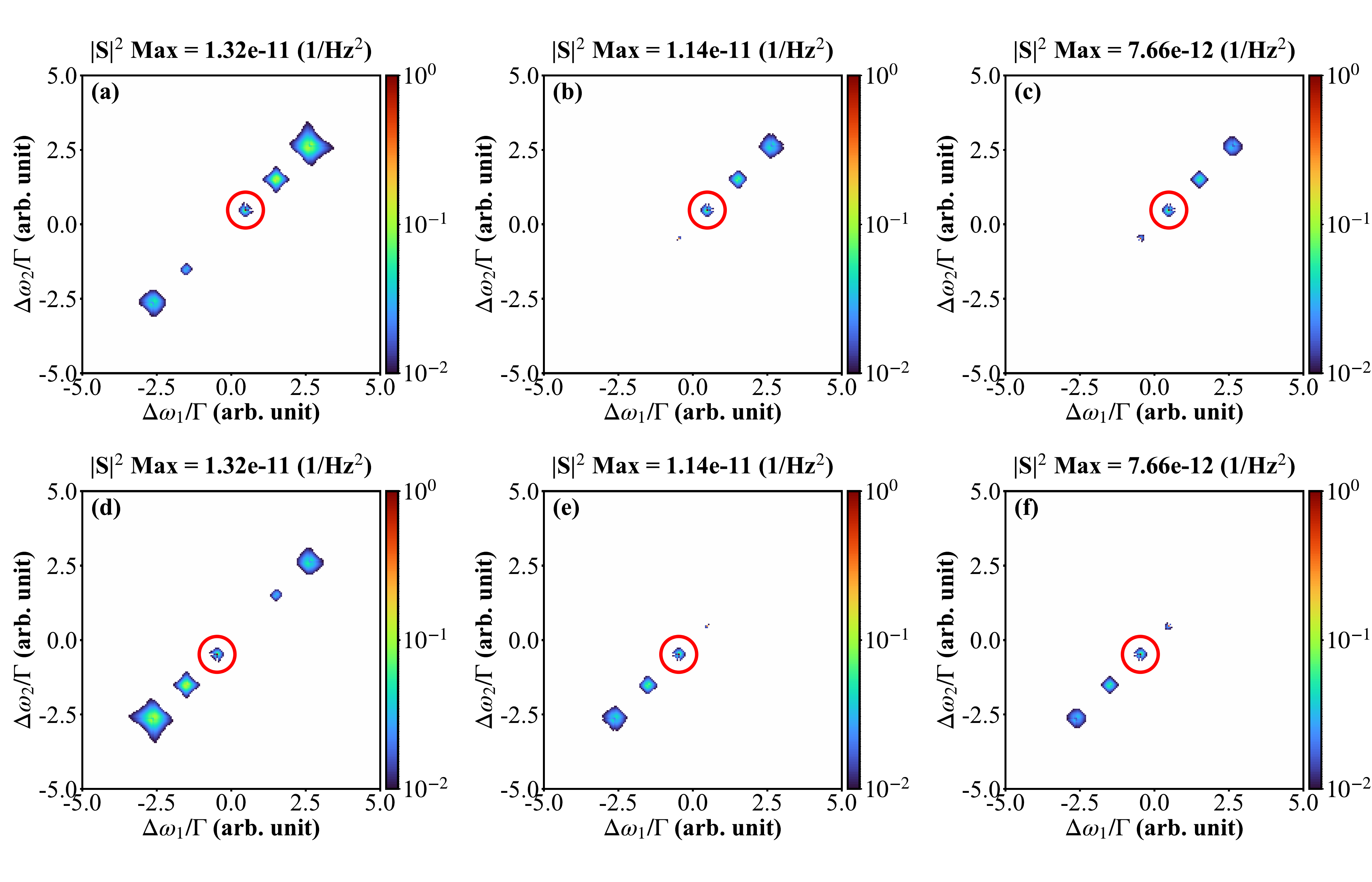}
    \caption{Inelastic scattering spectra for two incident photons to orthogonal model as a function of the probe detuning $\Delta\omega$ with $N = 2$ and $M = 10$. Panels (a)–(c) show the conventional EIT model; panels (d)–(f) show the orthogonal model. The red circles inside each graph indicate the detuning values at which the scattering magnitudes reach their maxima. The simulation parameters are $\Gamma_{1D} = \Gamma$, $\Gamma_e = 0.1\Gamma$, $J_1 = 3\Gamma$, $J_2 = 6\Gamma$, $\Delta_c = 0$, $k_0 d = \pi/2$, and $a = d$.}
    \label{Fig. Inelastic scattering inside Orthogonal model M = 10}
\end{figure}
\noindent
\begin{figure}[H]
    \centering
    \includegraphics[width=1.0\textwidth]{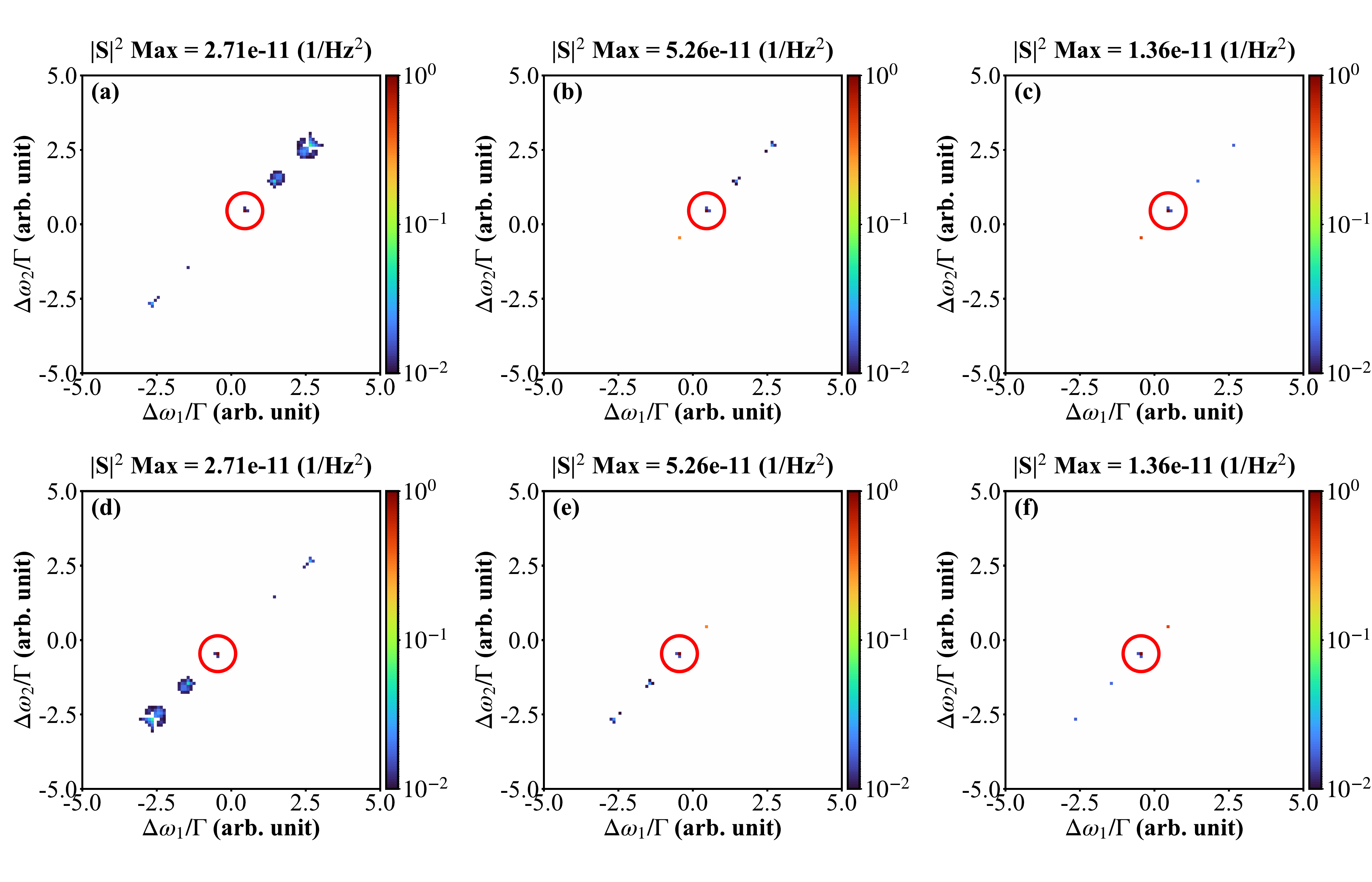}
    \caption{Inelastic scattering spectra for two incident photons to orthogonal model as a function of the probe detuning $\Delta\omega$ with $N = 2$ and $M = 20$. Panels (a)–(c) show the conventional EIT model; panels (d)–(f) show the orthogonal model. The red circles inside each graph indicate the detuning values at which the scattering magnitudes reach their maxima. The simulation parameters are $\Gamma_{1D} = \Gamma$, $\Gamma_e = 0.1\Gamma$, $J_1 = 3\Gamma$, $J_2 = 6\Gamma$, $\Delta_c = 0$, $k_0 d = \pi/2$, and $a = d$.}
    \label{Fig. Inelastic scattering inside Orthogonal model M = 20}
\end{figure}
\noindent
The inelastic scattering spectra in Figs. \ref{Fig. Inelastic scattering inside Orthogonal model M = 10} and \ref{Fig. Inelastic scattering inside Orthogonal model M = 20} reveal that four-wave mixing (FWM) in the orthogonal lattice is both spectrally selective and strongly enhanced. Nonlinear scattering is confined to a small number of well-localized modes within $[-5\Gamma,5\Gamma]$, while inelastic processes are strongly suppressed elsewhere, indicating that the structured reservoir effectively filters unwanted scattering channels. For $M=20$, the peak inelastic intensity reaches $5.26\times10^{-11}\,\mathrm{Hz}^{-1}$ at $\Delta\omega=\pm0.5\Gamma$, which is approximately six orders of magnitude larger than that of the normal configuration [see Fig. \ref{Fig. Inelastic scattering inside Zigzag model}(a-c)]. This enhancement originates from the constructive interplay between superradiant transmission resonances and subradiant absorption bands, which concentrates nonlinear coupling near selected frequencies without broadening the spectral response. Inside each inelastic mode, the orthogonal lattice also supports phase-stable nonlinear characteristics. As shown in Fig. \ref{Fig. B2} (Appendix \ref{sec: APPENDIX B}), the real part of the susceptibility $\chi'$ exhibits a piecewise-linear structure, interspersed with frequency intervals where $\chi'$ remains close to zero. This leads to a weakly frequency-dependent, or nearly vanishing, dispersive phase response, effectively decoupling dispersive phase accumulation among different localized nonlinear modes. Consequently, the accumulated phase of the generated idler photons becomes largely insensitive to frequency detuning, suppressing dispersive phase noise. This results in self-stabilized phases of individual nonlinear modes and a strong suppression of differential phase accumulation across the spectrum, enabling phase-robust four-wave mixing in waveguide QED, while remaining fundamentally distinct from active mode-locking mechanisms in conventional laser or frequency-comb systems.

\section {V.   CONCLUSIONS}
\addcontentsline{toc}{section}{CONCLUSIONS}
In this work, we introduced and analyzed two two-dimensional atomic lattice models coupled to a photonic crystal waveguide (PCW) to realize distinct EIT-based quantum functionalities: a zigzag lattice designed for quantum memory and an orthogonal lattice optimized for four-wave mixing (FWM). By engineering the lattice geometry through controlled interatomic spacing, inter-cell phase relations, and vertical positioning of atoms relative to the waveguide, the conventional EIT transparency window is reshaped into multiple transmission passbands separated by absorption bands. This design yields a structured transmission spectrum with enhanced spectral selectivity.\\
We showed that the resulting inelastic scattering is highly localized in frequency space, forming a small number of well-defined nonlinear modes. In the zigzag configuration, such spectral localization suppresses unwanted scattering channels and reduces crosstalk, improving the robustness of quantum memory operation. In the orthogonal configuration, the inelastic scattering intensity reaches $5.26\times10^{-11}\,\mathrm{Hz}^{-1}$ for $M=20$, which is approximately six orders of magnitude larger than that of the conventional lattice at $\Delta\omega=\pm0.5\Gamma$. This enhancement arises from the constructive interplay between superradiant transmission resonances and subradiant absorption bands, which concentrates nonlinear coupling near selected frequencies while preserving narrow spectral widths.\\
Our results highlight how lattice geometry, atom number, and detuning can be jointly exploited to control both the spectral distribution and efficiency of nonlinear interactions in waveguide QED systems. The ability to simultaneously engineer transparency windows, absorption bands, and phase-stable localized scattering modes establishes a versatile platform for on-demand photon generation, high-fidelity quantum storage, and tailored nonlinear optical responses. These findings open a pathway toward scalable integrated quantum photonic devices, where precise control of local light–matter interactions enables programmable quantum networks and spectrally selective inelastic scattering distribution.
\section {ACKNOWLEDGMENTS}
\label{sec: ACKNOWLEDGMENTS}
\addcontentsline{toc}{section}{ACKNOWLEDGMENTS}
This research is funded by the Vietnam National Foundation for Science and Technology Development (NAFOSTED) under grant Number \textbf{103.03.2025.142}.
\appendix
\section {APPENDIX A. THEORETICAL SUPPLEMENTARY}
\label{sec: APPENDIX A}
\addcontentsline{toc}{section}{Appendix A. Theoretical supplementary}
\subsection {A.1 Transmission via the Green’s-function formalism}
\label{sec: APPENDIX A1}
\addcontentsline{toc}{subsection}{A.1 Transmission via the Green’s-function formalism}
Scattering processes involving one or two photons in waveguide QED (wQED) systems can be compactly described using the scattering matrix ($S$-matrix) formalism. For a many-atom bidirectional waveguide, the single-photon transmission coefficient $T_k$ is defined through the $S$-matrix element $S_{p+;k+^{(1)} \equiv \bra{0} b_{+,out}(p)b_{+,in}(p)} \ket{0} \equiv T_k\delta_{p,k}$ where $b_{+,{\rm in(out)}}$ are the input (output) field operators for right-propagating modes, and the subscripts $\pm$ denote the propagation direction~\cite{Caneva_2015}. Using the input–output formalism and the single-excitation Green’s function $G_0$, the transmission coefficient can be expressed in terms of the approximately $N\times N$ Green’s function matrix $G_0$, whose elements are defined in the basis of atomic lowering and raising operators $\sigma$ and $\sigma^\dagger$, with $z_i$ denoting the position of the $i$-th atom~\cite{Caneva_2015}.
\begin{align}
T_k = 1 - \frac{\Gamma_{1D}^{i,j}}{2}\sum_{i,j}[G_0(k)]_{i,j}^{\sigma,\sigma^{\dagger}}e^{-ik_0(z_i-z_j)}.
\label{2}
\end{align}
The single-excitation Green’s function is given by $G_0 (\omega) = 1/(\omega - \mathcal{H}_1)$ where $\mathcal{H}_1$ is the effective non-Hermitian single-particle Hamiltonian in the rotating frame [Eq. \ref{Eq. Hamiltonian}]. The collective dissipative coupling into the waveguide is defined as $\Gamma_{1D}^{i,j}=\sqrt{\Gamma_{1D}^i\Gamma_{1D}^j}$.
\begin{align}
\mathcal{H}_1 =
\begin{pmatrix}
\left(\Delta_L- i\frac{\Gamma'}{2}\right)\delta_{ij}
- H_{1D} + H_{\mathcal{J}} & \Omega_c \delta_{ij} \\
\Omega_c \delta_{ij} & - 0
\end{pmatrix}.
\label{3}
\end{align}
with $\Delta_L$ denotes the detuning of probe field from control field.\\
The two-photon scattering matrix for photons incident from the left and propagating to the right consists of an elastic and an inelastic contribution~\cite{PhysRevA.93.013828,PhysRevLett.98.153003}. In this work, we focus exclusively on the inelastic component, which captures photon–photon correlations arising from nonlinear interactions mediated by the atomic ensemble and directly reflects the influence of lattice geometry on the spectral properties of the scattered photons. Since the detuning $\Delta\omega$ in the rotating frame is much smaller than the atomic resonance frequency $\omega_0$, frequency shifts imposed by energy conservation can be neglected. Under this approximation, the two-photon $S$-matrix is given by Eq. \ref{eq:S_matrix} \cite{Caneva_2015}.
\begin{flalign}
& S_{p_1+,p_2+,k_1+,k_2+}^{(2)} =
T_{k_1}T_{k_2}\delta_{p_1,k_1}\delta_{p_2,k_2}
- i\sum_{ij,i'j'}\sum_{\sigma_1\sigma_1';\sigma_2\sigma_2'}
\left[w^*(p_1,p_2)\right]_{ij}^{\sigma_1\sigma_1'}
[\mathbf{T}(E)]_{ij,i'j'}^{\sigma_1\sigma_1';\sigma_2\sigma_2'}
\left[w(k_1,k_2)\right]_{i'j'}^{\sigma_2\sigma_2'} & \\[3pt]
& \left[w(k_1,k_2)\right]_{ij}^{\sigma\sigma'} =
\sum_{j_1,j_2}\sqrt{\Gamma_{1D}^{ij_1}\Gamma_{1D}^{jj_2}}
\, e^{i(k_1 z_{j_1} + k_2 z_{j_2})}
\left[G_0(k_1)\right]_{i'j_1}^{\sigma\alpha}
\left[G_0(k_2)\right]_{i'j_2}^{\sigma'\alpha} &
\label{eq:S_matrix}
\end{flalign}
In this expression, $\mathbf{T}(E) = (\mathbf{H}_2 - E)$, with $E = \omega_1 + \omega_2$ and $\mathbf{H}_2 = \mathbf{H}_1 \otimes I_{2N} + I_{2N} \otimes \mathbf{H}_1$.
\subsection {A.2 Interactions in many-body systems}
\label{sec: APPENDIX A2}
\addcontentsline{toc}{subsection}{A.2 Interactions in many-body systems}
In both models, coherent dipole–dipole exchange interactions occur within each unit cell and between neighboring cells, described by the Hamiltonian term $H_{\mathcal{J}}$. These interactions arise from virtual photon exchange mediated by the photonic crystal waveguide and can be tuned using engineered coupling elements such as flux-tunable couplers or bus resonators~\cite{PhysRevApplied.19.064043,PhysRevX.13.031035,Majer2007,Stassi2020}. By controlling the interatomic spacing and emitter–waveguide separation (Figs. \ref{Fig. One-photon transmission through Zigzag model}, \ref{Fig. One photon transmission through Orthogonal model}), the resulting exchanges shape the bandwidth and structure of the EIT transparency window and enable collective photonic resonances. Furthermore, the inclusion of auxiliary atomic chains with vanishing waveguide decay rates ($\Gamma_{1D}=0$) introduces dark interference channels, leading to the splitting of a single EIT window into multiple hybridized transparency bands. The effective Hamiltonians governing these interactions are given in Eqs. (\ref{6}, \ref{7}) for Model~1 and Model~2, respectively \cite{PhysRevA.97.043848}.
\begin{align}
H_{J,M1} = \sum_{m}^{M}\sum_{i,j}^{2}\mathcal{J}_1(\sigma_{eg}^{i,m}\sigma_{ge}^{j,m})\delta(|i-j|,1) + \sum_{i,j}^{M}\mathcal{J}_2 (\sigma_{eg}^{2,i}\sigma_{ge}^{1,j})\delta\left( \left\lfloor \tfrac{i}{2} \right\rfloor - \left\lfloor \tfrac{j}{2} \right\rfloor , 0 \right)
\label{6}\\
H_{J,M2} = \sum_{m}^{M}\sum_{i,j}^{N}\mathcal{J}_1(\sigma_{eg}^{i,m}\sigma_{ge}^{j,m})\delta(|i-j|,1) + \sum_{i,j}^{M}\mathcal{J}_2(\sigma_{eg}^{1,i}\sigma_{ge}^{1,j})\delta\left( \left\lfloor \tfrac{i}{2} \right\rfloor - \left\lfloor \tfrac{j}{2} \right\rfloor , 0 \right)
\label{7}
\end{align}
The spatial arrangement of $\Lambda$-type atoms along the waveguide controls their emission into the guided mode via Bragg-phase-dependent interference, enabling selective enhancement or suppression of collective resonances~\cite{PhysRevA.110.053716}. This mechanism structures the spectrum into alternating transmission and absorption bands, with an isolated EIT window emerging between adjacent absorption features. In the two-dimensional geometries considered, the waveguide coupling depends exponentially on the atomic height $h_i$~\cite{Lee2025ExcitationLocalization,pnas.1603788113}. Alternating strongly ($\Gamma_{1D}$) and weakly ($0$) coupled atoms further engineer the transmission spectrum, as described by the effective Hamiltonians in Eqs. (\ref{8}, \ref{9}) \cite{shen2007strongly}.
\begin{align}
H_{\Gamma_{1D}, M1} = i\sum_{m,n}^{2,M}\sum_{u,v}^{2,M}\sqrt{\Gamma_{1D}^{m,n}\Gamma_{1D}^{u,v}}(\sigma_{eg}^{m,n}\sigma_{ge}^{u,v})e^{ik_0|z_{m,n} - z_{u,v}|}
\label{8}\\
H_{\Gamma_{1D}, M2} = i\sum_{m,n}^{M}\sqrt{\Gamma_{1D}^{1,m}\Gamma_{1D}^{1,n}}(\sigma_{eg}^{1,m}\sigma_{ge}^{1,n})e^{ik_0|z_{1,m} - z_{1,n}|}
\label{9}
\end{align}
\section {APPENDIX B. DISPERSIVE-PHASE ANALYSIS VIA THE RETARDED GREEN’S FUNCTION}
\label{sec: APPENDIX B}
\addcontentsline{toc}{section}{Appendix B. Superradiant and subradiant modes via Green’s-function analysis}
In this Appendix, we analyze the collective eigenmodes of the proposed atomic structures using the real part of the retarded Green’s function, which directly determines the linear susceptibility. The real part governs frequency shifts and dispersive phase responses. This Green’s-function analysis provides a unified interpretation of the transmission resonances and associated phase behavior, revealing a dispersively stabilized phase response and enhanced phase robustness.
\noindent
\begin{figure}[H]
    \centering
    \includegraphics[width=0.7\textwidth]{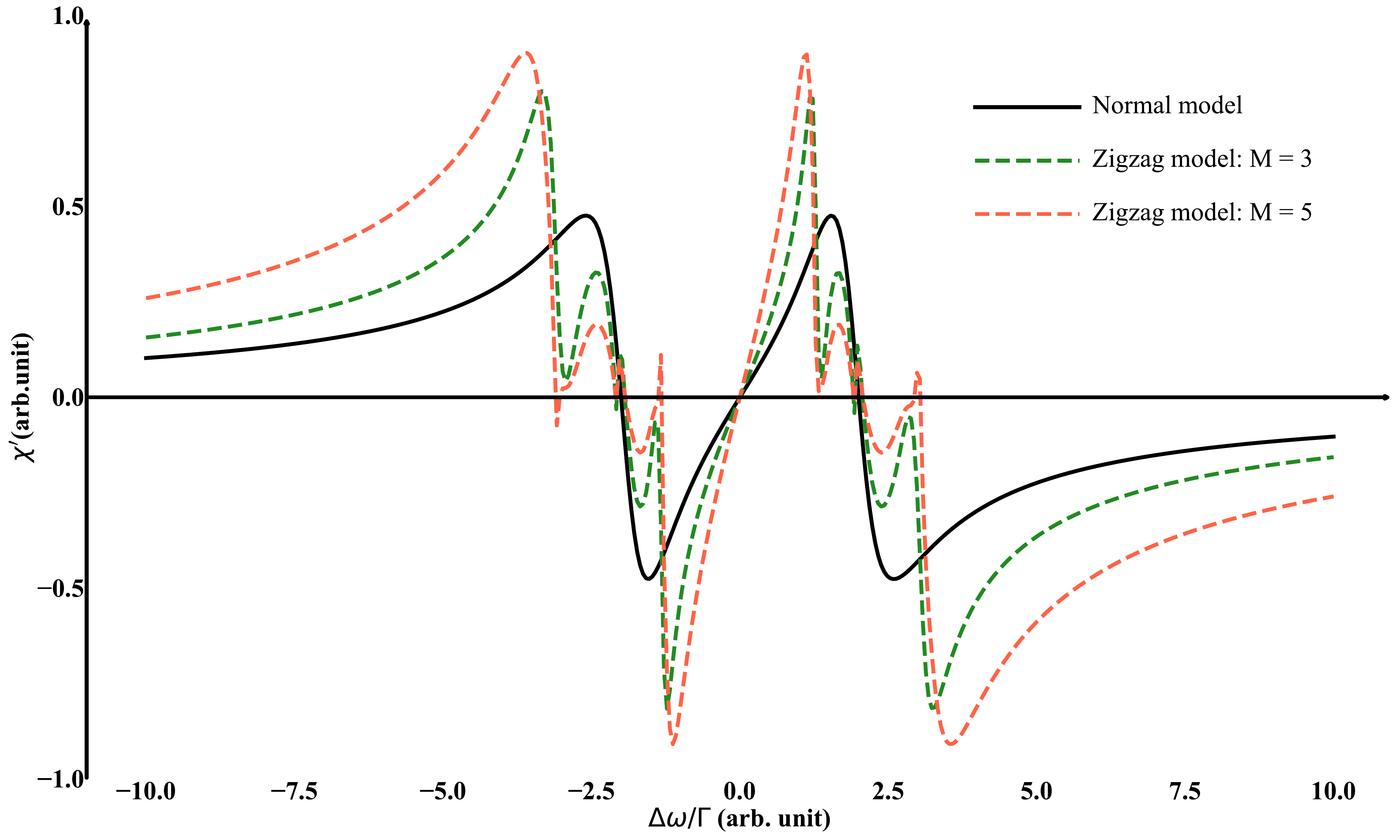}
    \caption{Real part $\chi'$ of the linear susceptibility for the zigzag $\Lambda$-atomic system atomic system as a function of dimensionless detuning $\Delta\omega/\Gamma$. The black solid line corresponds to the conventional EIT model, while the green and orange dashed lines represent the Zigzag model with two atoms per unit cell $N = 2$, and with $M=3$ and $M=5$, respectively. Simulation parameters are $\Gamma_{1D} = \Gamma$, $\Gamma_e = 0.1\Gamma$, $J_1 = 0.2\Gamma$, $J_2 = 1.6\Gamma$, $\Delta_c = 0$, $k_0 d = \pi/2$, and $a = d$.}
    \label{Fig. B1}
\end{figure}
\noindent
As shown in Fig. \ref{Fig. B1}, $\chi'$ approaches zero in narrow spectral regions adjacent to the EIT window. In these intervals, the dispersive phase response vanishes, indicating that transmission features at detunings $\Delta\omega/\Gamma \simeq \pm(2\text{--}3)$ are not governed by phase-induced interference but are dominated by absorption-related effects. The absence of linear dispersion in these regions can stabilize adiabatic evolution by suppressing unwanted phase accumulation.
\noindent
\begin{figure}[H]
    \centering
    \includegraphics[width=0.7\textwidth]{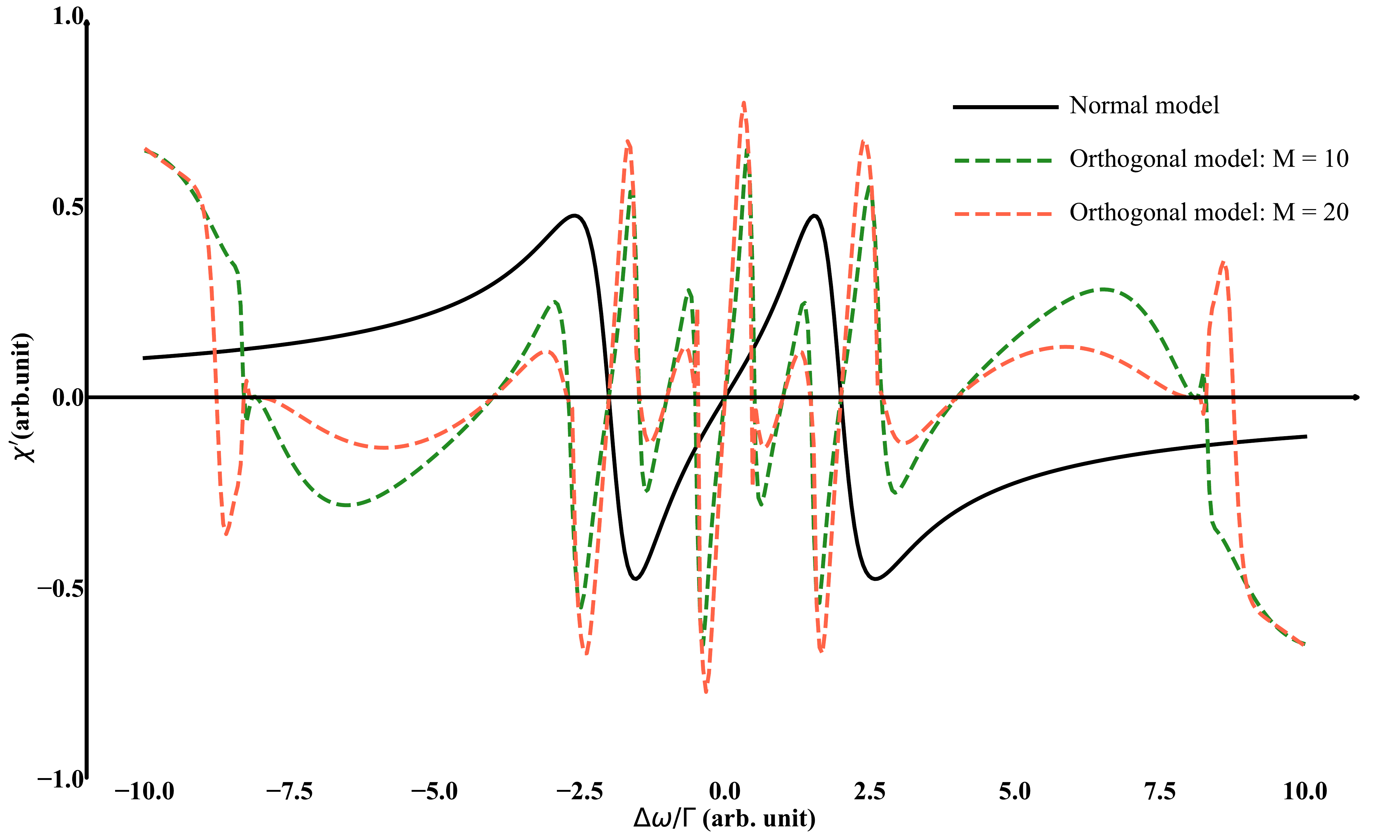}
    \caption{Real part $\chi'$ of the linear susceptibility for the orthogonal $\Lambda$-atomic system as a function of dimensionless detuning $\Delta\omega/\Gamma$. The black solid line corresponds to the conventional EIT model, while the green and orange dashed lines represent the Orthogonal models with two atoms per unit cell $N = 2$, and with $M=10$ (ten atomic pairs) and $M=20$ (twenty atomic pairs), respectively. Simulation parameters are $\Gamma_{1D} = \Gamma$, $\Gamma_e = 0.1\Gamma$, $J_1 = 3\Gamma$, $J_2 = 6\Gamma$, $\Delta_c = 0$, $k_0 d = \pi/2$, and $a = d$.}
    \label{Fig. B2}
\end{figure}
\noindent
For the orthogonal configuration (Fig. \ref{Fig. B2}), $\chi'$ exhibits multiple zero crossings at detunings $\Delta\omega/\Gamma \simeq \pm1$ and $\pm3$, which become more pronounced as the number of cells increases. At these frequencies, the vanishing dispersive response suppresses linear phase-matching contributions, implying that four-wave mixing is driven predominantly by nonlinear coupling mediated by collective atomic coherences. The non-dispersive character further reduces phase errors and preserves coherence in the nonlinear scattering processes.
\bibliographystyle{unsrt}   
\bibliography{references/Bibb}

@article{PhysRevA.97.043848,
  title = {Multiple transparency windows and Fano interferences induced by dipole-dipole couplings},
  author = {Diniz, E. C. and Borges, H. S. and Villas-Boas, C. J.},
  journal = {Phys. Rev. A},
  volume = {97},
  issue = {4},
  pages = {043848},
  numpages = {11},
  year = {2018},
  publisher = {American Physical Society},
}

@article{Caneva_2015,
year = {2015},
month = {oct},
publisher = {IOP Publishing},
volume = {17},
number = {11},
pages = {113001},
author = {Caneva, Tommaso and Manzoni, Marco T and Shi, Tao and Douglas, James S and Cirac, J Ignacio and Chang, Darrick E},
title = {Quantum dynamics of propagating photons with strong interactions: a generalized input–output formalism},
journal = {New Journal of Physics},
}

@article{pnas.1603788113,
author = {Jonathan D. Hood  and Akihisa Goban  and Ana Asenjo-Garcia  and Mingwu Lu  and Su-Peng Yu  and Darrick E. Chang  and H. J. Kimble },
title = {Atom–atom interactions around the band edge of a photonic crystal waveguide},
journal = {Proceedings of the National Academy of Sciences},
volume = {113},
number = {38},
pages = {10507-10512},
year = {2016},
}

@article{PhysRevA.98.023814,
  title = {Photon transport mediated by an atomic chain trapped along a photonic crystal waveguide},
  author = {Song, Guo-Zhu and Munro, Ewan and Nie, Wei and Kwek, Leong-Chuan and Deng, Fu-Guo and Long, Gui-Lu},
  journal = {Phys. Rev. A},
  volume = {98},
  issue = {2},
  pages = {023814},
  numpages = {9},
  year = {2018},
  month = {Aug},
  publisher = {American Physical Society},
}

@article{PhysRevA.110.053716,
  title = {Engineering photonic band gaps with a waveguide-QED structure containing an atom-polymer array},
  author = {Wang, M. S. and Jia, W. Z.},
  journal = {Phys. Rev. A},
  volume = {110},
  issue = {5},
  pages = {053716},
  numpages = {14},
  year = {2024},
  month = {Nov},
  publisher = {American Physical Society},
}

@article{Douglas2015,
author={Douglas, J. S.
and Habibian, H.
and Hung, C.-L.
and Gorshkov, A. V.
and Kimble, H. J.
and Chang, D. E.},
title={Quantum many-body models with cold atoms coupled to photonic crystals},
journal={Nature Photonics},
year={2015},
month={May},
day={01},
volume={9},
number={5},
pages={326-331}
}

@article{Baba2008,
author={Baba, Toshihiko},
title={Slow light in photonic crystals},
journal={Nature Photonics},
year={2008},
month={Aug},
day={01},
volume={2},
number={8},
pages={465-473},
issn={1749-4893}
}

@Article{Novikov2016,
author={Novikov, S.
and Sweeney, T.
and Robinson, J. E.
and Premaratne, S. P.
and Suri, B.
and Wellstood, F. C.
and Palmer, B. S.},
title={Raman coherence in a circuit quantum electrodynamics lambda system},
journal={Nature Physics},
year={2016},
month={Jan},
day={01},
volume={12},
number={1},
pages={75-79},
}

@article{Lee2025ExcitationLocalization,
  title        = {Controlling Excitation Localization in Waveguide QED Systems},
  author       = {Lee, C.-Y. and Lin, K.-T. and Lin, G.-D. and Jen, H. H.},
  journal      = {arXiv preprint arXiv:2505.20878},
  year         = {2025},
  eprint       = {2505.20878},
  archivePrefix= {arXiv},
  primaryClass = {quant-ph}
}

@article{PhysRevApplied.19.064043,
  title = {Modular Tunable Coupler for Superconducting Circuits},
  author = {Campbell, Daniel L. and Kamal, Archana and Ranzani, Leonardo and Senatore, Michael and LaHaye, Matthew D.},
  journal = {Phys. Rev. Appl.},
  volume = {19},
  issue = {6},
  pages = {064043},
  numpages = {17},
  year = {2023},
  month = {Jun},
  publisher = {American Physical Society}
}

@article{PhysRevX.13.031035,
  title = {High-Fidelity, Frequency-Flexible Two-Qubit Fluxonium Gates with a Transmon Coupler},
  author = {Ding, Leon and Hays, Max and Sung, Youngkyu and Kannan, Bharath and An, Junyoung and Di Paolo, Agustin and Karamlou, Amir H. and Hazard, Thomas M. and Azar, Kate and Kim, David K. and Niedzielski, Bethany M. and Melville, Alexander and Schwartz, Mollie E. and Yoder, Jonilyn L. and Orlando, Terry P. and Gustavsson, Simon and Grover, Jeffrey A. and Serniak, Kyle and Oliver, William D.},
  journal = {Phys. Rev. X},
  volume = {13},
  issue = {3},
  pages = {031035},
  numpages = {24},
  year = {2023},
  month = {Sep},
  publisher = {American Physical Society}
}

@Article{Majer2007,
author={Majer, J.
and Chow, J. M.
and Gambetta, J. M.
and Koch, Jens
and Johnson, B. R.
and Schreier, J. A.
and Frunzio, L.
and Schuster, D. I.
and Houck, A. A.
and Wallraff, A.
and Blais, A.
and Devoret, M. H.
and Girvin, S. M.
and Schoelkopf, R. J.},
title={Coupling superconducting qubits via a cavity bus},
journal={Nature},
year={2007},
month={Sep},
day={01},
volume={449},
number={7161},
pages={443-447}
}

@Article{Stassi2020,
author={Stassi, Roberto
and Cirio, Mauro
and Nori, Franco},
title={Scalable quantum computer with superconducting circuits in the ultrastrong coupling regime},
journal={npj Quantum Information},
year={2020},
month={Aug},
day={06},
volume={6},
number={1},
pages={67}
}

@article{tevcer2024strongly,
  title={Strongly interacting photons in 2D waveguide QED},
  author={Te{\v{c}}er, Matija and Di Liberto, Marco and Silvi, Pietro and Montangero, Simone and Romanato, Filippo and Calaj{\'o}, Giuseppe},
  journal={Physical Review Letters},
  volume={132},
  number={16},
  pages={163602},
  year={2024},
  publisher={APS}
}

@article{RevModPhys.77.633,
  title = {Electromagnetically induced transparency: Optics in coherent media},
  author = {Fleischhauer, Michael and Imamoglu, Atac and Marangos, Jonathan P.},
  journal = {Rev. Mod. Phys.},
  volume = {77},
  issue = {2},
  pages = {633--673},
  numpages = {0},
  year = {2005},
  month = {Jul},
  publisher = {American Physical Society},
}

@article{PhysRevA.104.013705,
  title = {Optimal two-photon excitation of bound states in non-Markovian waveguide QED},
  author = {Trivedi, Rahul and Malz, Daniel and Sun, Shuo and Fan, Shanhui and Vu\ifmmode \check{c}\else \v{c}\fi{}kovi\ifmmode \acute{c}\else \'{c}\fi{}, Jelena},
  journal = {Phys. Rev. A},
  volume = {104},
  issue = {1},
  pages = {013705},
  numpages = {9},
  year = {2021},
  month = {Jul},
  publisher = {American Physical Society},
}

@article{PhysRevLett.131.193603,
  title = {Zeno Regime of Collective Emission: Non-Markovianity beyond Retardation},
  author = {Zhang, Yu-Xiang},
  journal = {Phys. Rev. Lett.},
  volume = {131},
  issue = {19},
  pages = {193603},
  numpages = {7},
  year = {2023},
  month = {Nov},
  publisher = {American Physical Society},
}

@Article{Ding2013,
author={Ding, Dong-Sheng
and Zhou, Zhi-Yuan
and Shi, Bao-Sen
and Guo, Guang-Can},
title={Single-photon-level quantum image memory based on cold atomic ensembles},
journal={Nature Communications},
year={2013},
month={Oct},
day={02},
volume={4},
number={1},
pages={2527}
}

@article{PhysRevA.102.063720,
  title = {Broadband coherent optical memory based on electromagnetically induced transparency},
  author = {Wei, Yan-Cheng and Wu, Bo-Han and Hsiao, Ya-Fen and Tsai, Pin-Ju and Chen, Ying-Cheng},
  journal = {Phys. Rev. A},
  volume = {102},
  issue = {6},
  pages = {063720},
  numpages = {11},
  year = {2020},
  month = {Dec},
  publisher = {American Physical Society}
}

@article{PhysRevA.110.063723,
  title = {Frequency-tunable biphoton generation via spontaneous four-wave mixing},
  author = {Shiu, Jiun-Shiuan and Lin, Chang-Wei and Huang, Yu-Chiao and Lin, Meng-Jung and Huang, I-Chia and Wu, Ting-Ho and Kuan, Pei-Chen and Chen, Yong-Fan},
  journal = {Phys. Rev. A},
  volume = {110},
  issue = {6},
  pages = {063723},
  numpages = {8},
  year = {2024},
  month = {Dec},
  publisher = {American Physical Society}
}

@article{PhysRevApplied.14.024013,
  title = {Impact of Four-Wave-Mixing Noise from Dense Wavelength-Division-Multiplexing Systems on Entangled-State Continuous-Variable Quantum key Distribution},
  author = {Du, Shanna and Tian, Yan and Li, Yongmin},
  journal = {Phys. Rev. Appl.},
  volume = {14},
  issue = {2},
  pages = {024013},
  numpages = {12},
  year = {2020},
  month = {Aug},
  publisher = {American Physical Society}
}

@article{Prajapati:17,
author = {Nikunj Prajapati and Gleb Romanov and Irina Novikova},
journal = {J. Opt. Soc. Am. B},
number = {9},
pages = {1994--1999},
publisher = {Optica Publishing Group},
title = {Suppression of four-wave mixing in hot rubidium vapor using ladder scheme Raman absorption},
volume = {34},
month = {Sep},
year = {2017},
}

@article{PhysRevB.96.144201,
  title = {Two mechanisms of disorder-induced localization in photonic-crystal waveguides},
  author = {Garc\'{\i}a, P. D. and Kir\ifmmode \check{s}\else \v{s}\fi{}ansk\ifmmode \dot{e}\else \.{e}\fi{}, G. and Javadi, A. and Stobbe, S. and Lodahl, P.},
  journal = {Phys. Rev. B},
  volume = {96},
  issue = {14},
  pages = {144201},
  numpages = {5},
  year = {2017},
  month = {Oct},
  publisher = {American Physical Society}
}

@article{PhysRevA.93.013828,
  title = {Green's-function formalism for waveguide QED applications},
  author = {Schneider, Michael P. and Sproll, Tobias and Stawiarski, Christina and Schmitteckert, Peter and Busch, Kurt},
  journal = {Phys. Rev. A},
  volume = {93},
  issue = {1},
  pages = {013828},
  numpages = {17},
  year = {2016},
  month = {Jan},
  publisher = {American Physical Society}
}

@article{PhysRevLett.98.153003,
  title = {Strongly Correlated Two-Photon Transport in a One-Dimensional Waveguide Coupled to a Two-Level System},
  author = {Shen, Jung-Tsung and Fan, Shanhui},
  journal = {Phys. Rev. Lett.},
  volume = {98},
  issue = {15},
  pages = {153003},
  numpages = {4},
  year = {2007},
  month = {Apr},
  publisher = {American Physical Society}
}

@Article{Offer2018,
author={Offer, R. F.
and Stulga, D.
and Riis, E.
and Franke-Arnold, S.
and Arnold, A. S.},
title={Spiral bandwidth of four-wave mixing in Rb vapour},
journal={Communications Physics},
year={2018},
month={Nov},
day={21},
volume={1},
number={1},
pages={84},
issn={2399-3650}
}

@article{RevModPhys.89.015006,
  title = {Stimulated Raman adiabatic passage in physics, chemistry, and beyond},
  author = {Vitanov, Nikolay V. and Rangelov, Andon A. and Shore, Bruce W. and Bergmann, Klaas},
  journal = {Rev. Mod. Phys.},
  volume = {89},
  issue = {1},
  pages = {015006},
  numpages = {66},
  year = {2017},
  month = {Mar},
  publisher = {American Physical Society}
}

@article{shen2007strongly,
  title={Strongly Correlated Two-Photon Transport in a One-Dimensional Waveguide Coupled to a Two-Level System},
  author={Shen, Jung-Tsung and Fan, Shanhui},
  journal={Physical review letters},
  volume={98},
  number={15},
  pages={153003},
  year={2007},
  publisher={APS}
}
\end{document}